\documentclass[twocolumn,aps,showpacs,amsmath]{revtex4}
\usepackage{dcolumn}
\usepackage{bm}
\begin{document}

\title{Minimal 3-3-1 model, lepton mixing
and muonium-antimuonium conversion}
\author{A. Gusso}
\email{gusso@ift.unesp.br}
\affiliation{Instituto de  F\'{\i}sica Te\'{o}rica, Universidade
Estadual Paulista, Rua Pamplona 145,
01405-900 S\~{a}o Paulo - SP, Brazil}
\author{C. A de S. Pires}
\email{cpires@fisica.ufpb.br}
 \affiliation{Departamento de F\'{\i}sica, Universidade Federal da
Para\'{\i}ba, Caixa Postal 5008, 58051-970, Jo\~ao Pessoa - PB,
Brazil.}
\author{P. S. Rodrigues da Silva}
 \email{fedel@ift.unesp.br}
\affiliation{Instituto de  F\'{\i}sica Te\'{o}rica, Universidade
Estadual Paulista, Rua Pamplona 145,
01405-900 S\~{a}o Paulo - SP, Brazil}

\date{\today}

\begin{abstract}
The recent experimental results on neutrino oscillation and on
muonium-antimuonium conversion require extension of the minimal
3-3-1 model. We review the constraints imposed to the model by
those measurements and suggest a pattern of leptonic mixing, with
charged leptons in a non-diagonal basis, which accounts for the
neutrino physics and circumvents the tight muonium-antimuonium
bounds on the model. We also illustrate a scenario where this
pattern could be realized.
\end{abstract}

\maketitle

\section{Introduction}

The  minimal 3-3-1 \cite{viceframp} model is an alternative to
the standard model (SM) of the electro-weak interactions that,
among other noteworthy features, presents an interesting leptonic
phenomenology due to the presence of bileptons. The  charged
currents involving the vector bileptons $V^{\pm}$ and
$U^{\pm\pm}$  allow for a number of rare  processes, among them is
the muonium-antimuonium($M-\overline{M}$) conversion.

Presently, there are two sources of experimental constraints
claiming an extension of the minimal model. One of them stems
from the recent experimental results that corroborate the
neutrino oscillation hypothesis~\cite{SNO,K2K,KamLAND,BahcallConcha}.

In this regards, it was shown in Ref.~\cite{joshi} that the model
disposes of a potential that leads to  a type II seesaw mechanism
when we consider terms that break explicitly the lepton number.
This means that the model leads naturally to small neutrino mass,
as usually required by the neutrino oscillation hypothesis.
However small neutrino mass is not the whole issue in neutrino
physics once neutrino oscillation requires neutrino mixing. It
was shown in Ref.~\cite{gusso} that the minimal model is unable of
generating  the pattern of neutrino mixing required by solar and
atmospheric neutrino oscillation\cite{gusso}.

The other source of  constraints on the model comes from the
limits imposed on the $M-\overline{M}$ conversion. In
Ref.~\cite{muoexp}, it was posed that such a constraint implies a
lower bound on the doubly charged vector bilepton in the minimal
3-3-1 model, $ M_{U^{++}} \geq 850 $ GeV, which is in clear
contradiction to the predicted upper bound required by
self-consistency of the model, namely, $M_{U^{++}} \leq 600 $
GeV~\cite{mu++limit}.

As we have two different sources of constraints implying that for
the minimal 3-3-1 model to be viable it has to be extended, and
both involve its leptonic sector, it seems interesting to look for
extensions that could lead to the desired pattern of neutrino
mixing and simultaneously get rid of the $M-\overline{M}$ bounds.
This is the aim of this work, and we shall carry on this proposal
dealing with the leptonic mixing.

This work is organized as follows. In section  \ref{sec1} we
present the leptonic sector of the minimal 3-3-1 model,
highlighting the aspects that are relevant to our analysis. In
section \ref{sec2}, we show how it is possible to simultaneously
be consistent with neutrino data and circumvent the bound imposed
by the non-observation of $M-\overline{M}$ conversion setting an
appropriate mixing in the leptonic sector. We also present the
texture of the leptonic mass matrices that would lead to such a
mixing and suggest a suitable way in which this scenario can be
realized, addressing its phenomenological situation. We finally
conclude in section \ref{sec3}.

\section{Leptonic sector  of the minimal 3-3-1 model}
\label{sec1}

Before addressing  the main point of our investigation, it is
worthwhile to review how the lepton masses are generated in the
minimal 3-3-1 model. The Yukawa interactions that lead to these
masses are given by,
\begin{eqnarray}
{\cal L}^Y_l&=&\frac{G_{ab}}{2}\overline{ (\Psi_{aL})^c}
S^*\Psi_{b_L}+ \frac{F_{ab}}{2}F_{ab}\epsilon^{i j k}
\overline{(\Psi_{iaL})^c} \Psi_{jbL} \eta^*_k \nonumber \\
&&+ \mbox{H.c},
\label{yukawa} \end{eqnarray}
where $a,\,b$ label the different families and $i,\,j,\,k =
1,2,3$ label the elements of each multiplet, with $\Psi_{aL} =(
\nu_a , e_a , e^c_a)_L^T $,  $ \eta =( \eta^0 , \eta^-_1 ,
\eta^+_2)^T $  and the sextet $S$,
\begin{eqnarray}
S=\left(\begin{array}{ccc}
 \sigma^0_1 & h_1^{-} & s_2^{+} \\
 h_1^{-} & H_1^{--} & \sigma^0_2 \\
 h_2^{+} & \sigma^0_2 & H_2^{++}
\end{array}
\right).
 \label{sextet}
 \end{eqnarray}

In general non-diagonal matrices  $M_l$  and $M_\nu$  can be diagonalized as follows,
\begin{eqnarray} M_l^D = V_{eL}^\dag M_lV_{eR}\,,\,\,\,\,\,\, M_\nu^D =
V_{\nu}^\dag M_\nu V_{\nu}\,, \label{diagmass} \end{eqnarray}
where $M_{l,\nu}$ is the mass matrix in the interaction basis, and
$M^D_{l,\nu}$ is the mass matrix in the mass basis. The matrices
$V$ transform the lepton fields from the interaction eigenstates
into mass eigenstates and, in principle, they are different for
left-handed and right-handed fields. These diagonalization
matrices combine themselves in the charged current of the model
which, after the 3-3-1 breaking to the $SU(3)_c \otimes
U(1)_{EM}$, is given by (omitting family indices):
\begin{eqnarray}
{\cal L}^{CC}_{l}&=& -\frac{g}{\sqrt{2}} \overline{
e_L}
\gamma^\mu O^W\nu_LW^-_\mu -\frac{g}{\sqrt{2}}\overline{(e_R)^c}O^V\gamma^\mu
\nu_L V^-_\mu \nonumber \\
&& -\frac{g}{\sqrt{2}}\overline{(e_R)^c}O^U \gamma^\mu
e_LU^{--}_\mu + \mbox{H.c.}\,,
\label{cc}
\end{eqnarray}
where $O^W=V^T_{eL} V_{\nu} $, $O^V=V^T_{eR} V_{\nu}$, and
$O^U=V^T_{eR} V_{eL}$ are the mixing matrices. While the
experimental data concerning neutrino oscillation give information
on the possible values for the elements of $O^W$, we have no such
information on the elements of $O^V$ and $O^U$. We also note that
since the right handed mixing matrix $V_{eR}$ enters into the
charged current interactions through bilepton-lepton coupling, the
leptonic mixing results more complex than we would expect in
simple extensions of the SM.

The mixing matrices appearing in this work are all of the
Maki-Nakagawa-Sakata (MNS) type\cite{MNS} and we will be
considering the simplest case of a zero CP violating phase, since
this phase is irrelevant throughout our analysis. This implies
that these matrices are real, and the hermitian conjugate of the
matrices involved are simply their transpose. We then adopt the
following parameterization for these matrices $O^{W\,,V\,,U}$,
\begin{eqnarray}{\small
\left (
\begin{array}{ccccc}
c_{12} c_{13} &\,\,& s_{12} c_{13} &\,\,& s_{13}  \\
-s_{12} c_{23}- c_{12} s_{13} s_{23} &\,\,& c_{12} c_{23}-s_{12}
s_{13} s_{23} &\,\,&
c_{13} s_{23} \\
s_{12} s_{23} -c_{12}s_{13}c_{23}&\,\,& -c_{12} s_{23}-s_{12}
s_{13} c_{23} &\,\,& c_{13} c_{23}
\end{array}
\right )}, \nonumber \\ \label{MNS}
\end{eqnarray}
where, as usual, $s_{ij}$ and $c_{ij}$ denote sines and cosines
of their arguments $\theta_{ij}$. This parameterization will be
used in section~\ref{sec2} when we discuss the appropriate
pattern for lepton mixing which would allow us to get rid of the
$M-\overline{M}$ bound, and still be consistent with neutrino
physics.

\section{Could $M-\overline{M}$ conversion be absent in 3-3-1 model?}
\label{sec2}

As we saw in section \ref{sec1} the leptonic sector of the 3-3-1
model presents three mixing matrices $O^W$, $O^V$, and $O^U$. It
is opportune to observe that these mixing matrices are not
completely independent of each other, since they involve products
of the lepton diagonalization matrices $V_\nu$, $V_{eL}$, and
$V_{eR}$. In the case of a diagonal charged lepton basis $V_{eL}$
and $V_{eR}$ are also diagonal, leading to a diagonal mixing
matrix for the interactions among $U^{\pm \pm}$ and the leptons.
Hence, as we want to find a way of overcoming the $M-\overline{M}$
bound by fixing an appropriate form for the matrix $O^U$, we must
consider the leptonic sector in a non-diagonal charged lepton
basis. Then, our first task is to determine the three matrices
$V_\nu$, $V_{eL}$ and $V_{eR}$. The data from neutrino physics
provide certain knowledge about the mixing matrix $O^W$. From it
we can infer $V_\nu$ and $V_{eL}$. However we do not dispose of
sufficient data to determine $V_{eR}$ and this gives us some room
to make a key assumption in this work, namely, imposing that
$V_{eR}$ mimics $V_{eL}$. The reason behind such an assumption is
that it realizes our proposal of circumventing the
$M-\overline{M}$ bounds in the 3-3-1 model in a neat way, as we
expose next. We then finish this section by commenting about the
phenomenological status of this scenario.

Let us first determine the pattern of $O^W$. The recent analysis
of atmospheric neutrino still favors $\nu_\mu - \nu_\tau$
oscillation with an almost maximal mixing $0.92 <
\sin^22\theta_{atm} \leq 1.0$ at 90 \% C.L.~\cite{atmospheric}. We
also have that the oscillation among $\nu_e -\nu_\mu$ is almost
settled as the explanation for the  solar neutrino problem. Here
the recent results allow $0.25 \leq \sin^2 \theta_{sun}\leq
0.40$  and $0.6 \leq \cos^2 \theta_{sun}\leq
0.75$ (90 \% C.L.)~\cite{solar}. Since the CHOOZ experiment failed
to see the disappearance of $\bar \nu_e$, we also have $0\leq
\sin^2 2\theta_{chz} <0.1$ (90 \% C.L.)~\cite{chooz}. For our
proposal, we can fix the angles $\theta_{atm}$ ($\theta_{23}$)
and $\theta_{chz}$ ( $\theta_{13}$), which is straightforwardly
done by taking the
best fit for $\theta_{atm} =45^{\circ}$, while
$\theta_{chz}=0^{\circ}$, in agreement with the above presented
results. The angle involved in the solar neutrino oscillation
($\theta_{12}$) is the one that allows for a certain range of
values, and it can be kept as a free parameter in our
investigation. We are left then with the so called maximal mixing
pattern for $O^W$,
\begin{eqnarray} O^W= \left (
\begin{array}{ccc}
c & s & 0 \\
\frac{-s}{\sqrt{2}} & \frac{c}{\sqrt{2}} & \frac{1}{\sqrt{2}} \\
 \frac{s}{\sqrt{2}}& -\frac{c}{\sqrt{2}}  & \frac{1}{\sqrt{2}}
\end{array}
\right )\,, \label{neutmix} \end{eqnarray}
where we have used the short form $\sin \theta_{sun}=s$ and $\cos
\theta_{sun}=c$.

A non-diagonal charged lepton mass basis means that both, $V_{eL}$
and $V_\nu$, are non trivial. The only way of separating
Eq.~(\ref{neutmix}) in these two matrices is to have maximal
mixing between $\nu_\mu$ and $\nu_\tau$ coming from the charged
lepton sector and the mixing in the $\nu_e$ to $\nu_\mu$
oscillation coming from the neutrino sector. In order to
disentangle the contributions of the distinct diagonalization
matrices to $O^W$ given by Eq.~(\ref{neutmix}), we dissociate it
as:
\begin{eqnarray} O^W&&= \left (
\begin{array}{ccc}
c & s & 0 \\
\frac{-s}{\sqrt{2}} & \frac{c}{\sqrt{2}} & \frac{1}{\sqrt{2}}
\\
 \frac{s}{\sqrt{2}}& -\frac{c}{\sqrt{2}}  & \frac{1}{\sqrt{2}}
\end{array}
\right )\nonumber \\
&&=\left (
\begin{array}{ccc}
1 & 0 & 0 \\
0 & \frac{1}{\sqrt{2}} & \frac{1}{\sqrt{2}} \\
 0& -\frac{1}{\sqrt{2}}  & \frac{1}{\sqrt{2}}
\end{array}
\right )\times \left (
\begin{array}{ccc}
 c &  s  & 0 \\
 -s &  c  & 0 \\
 0 &  0  & 1
\end{array}
\right )\,, \label{division} \end{eqnarray}
which is the only way of doing it since otherwise we would
inevitably get a large $\theta_{chz}$. From this matrix equation
we can easily recognize the contribution from the charged lepton
sector (remembering that $O^W=V^T_{eL}V_\nu$),
\begin{eqnarray} V_{eL} =\left (
\begin{array}{ccc}
1 & 0 & 0 \\
0 & \frac{1}{\sqrt{2}} &- \frac{1}{\sqrt{2}} \\
 0& \frac{1}{\sqrt{2}}  & \frac{1}{\sqrt{2}}
\end{array}
\right ),
\label{clepton}
\end{eqnarray}
and the contribution from the neutrino sector,
\begin{eqnarray}
 V_{\nu}= \left (
\begin{array}{ccc}
c & s & 0 \\
-s & c & 0 \\
 0& 0  & 1
\end{array}
\right )\,. \label{neutpart} \end{eqnarray}
In this way the neutrino sector gets responsible for the mixing
related to the solar neutrino oscillation, and the charged lepton
sector to the maximal mixing related to the atmospheric neutrino
oscillation.

Let us take $O^V=V^T_{eR} V_{\nu}$ and dissociate it as the
product of three rotation matrices:
\begin{eqnarray} O^V&&= \left (
\begin{array}{ccc}
1 & 0 & 0  \\
0 &  c_{23} & s_{23}  \\
0 & -s_{23} & c_{23}
\end{array}
\right )\times \left (
\begin{array}{ccc}
c_{13} & 0 & s_{13}  \\
0 &  1 & 0  \\
-s_{13} & 0 & c_{13}
\end{array}
\right ) \nonumber \\
&&\times \left (
\begin{array}{ccc}
c_{12} & s_{12} & 0  \\
-s_{12} &  c_{12} & 0 \\
0 & 0 & 1
\end{array}
\right ).
\label{VeR1}
\end{eqnarray}

It is easy to recognize the last matrix at the right hand side of
Eq.~(\ref{VeR1}) as the neutrino mixing matrix given in
Eq.~(\ref{neutpart}). In this way let us take the usual notation
$c_{12}=c$ and $s_{12}=s$. The other two matrices must form the
$V_{eR}$ mixing matrix,
\begin{eqnarray} V^T_{eR}&&= \left (
\begin{array}{ccc}
1 & 0 & 0  \\
0 &  c_{23} & s_{23}  \\
0 & -s_{23} & c_{23}
\end{array}
\right )\times \left (
\begin{array}{ccc}
c_{13} & 0 & s_{13}  \\
0 &  1 & 0  \\
-s_{13} & 0 & c_{13}
\end{array}
\right ) \nonumber \\
&&=\left (
\begin{array}{ccc}
c_{13} & 0 & s_{13}  \\
-s_{23}s_{13} &  c_{23} & s_{23}c_{13}  \\
-c_{23}s_{13} & -s_{23} & c_{23}c_{13}
\end{array}
\right )\,. \label{VeR2} \end{eqnarray}
Since there is no information about the angles in the matrix
$V_{eR}$ we can, for convenience, assume that the charged current
mediated by $V^{\pm}$ does not contrast with the physics
experimentally established by the charged current mediated by
$W^{\pm}$. For that reason we adopt in Eq. (\ref{VeR2})
$\theta_{13}=0$, and $\theta_{23}$ maximal. This choice
accomplishes the goal of delivering the 3-3-1 model out of the
$M-\overline{M}$ bound if we require that the maximal angle be
negative, that is, $\theta_{23}=-45^{\circ}$. With these
assumptions we get
\begin{eqnarray} V_{eR} =\left (
\begin{array}{ccc}
1 & 0 & 0 \\
0 & \frac{1}{\sqrt{2}} & \frac{1}{\sqrt{2}} \\
 0&- \frac{1}{\sqrt{2}}  & \frac{1}{\sqrt{2}}
\end{array}
\right ).
\label{VeR}
\end{eqnarray}

The matrices in Eqs.~(\ref{clepton}), (\ref{neutpart}), and
(\ref{VeR}) determine the leptonic mixing in the 3-3-1 model.
$O^W$ is given by Eq.~(\ref{neutmix}) and the other two are
\begin{eqnarray} &&O^U=V^T_{eR}V_{eL}=\left (
\begin{array}{ccc}
1 & 0 & 0 \\
0 & 0 & -1 \\
0 & 1  & 0
\end{array}
\right ),\nonumber \\
&& O^V=V^T_{eL}V_\nu =\left (
\begin{array}{ccc}
c & s & 0 \\
-\frac{s}{\sqrt{2}} & \frac{c}{\sqrt{2}} & -\frac{1}{\sqrt{2}}
\\
\frac{s}{\sqrt{2}} & -\frac{c}{\sqrt{2}}  & \frac{1}{\sqrt{2}}
\end{array}
\right )\,.\label{VUmatrix} \end{eqnarray}

The ${M-\overline{M}}$ conversion is proportional to the product
of $O^U_{ee}$ and $O^U_{\mu \mu}$. According to the pattern of
leptonic mixing given above we have $O^U_{\mu \mu}=0$, which
completely eliminates the ${M-\overline{M}}$ conversion from the
3-3-1 model at tree level and, consequently, the bound on the
vector bilepton mass.

Finally, once we have the pattern of leptonic mixing, it is
imperative to establish the textures of the lepton mass matrices
that lead to such a pattern. With $V_\nu$, $V_{eR}$ and $V_{eL}$
given by Eqs.~ (\ref{clepton}), (\ref{neutpart}) and (\ref{VeR}),
respectively, we are able to obtain the textures of the neutrino
and charged lepton mass matrices through Eq.~(\ref{diagmass}).

Taking $M^D_\nu=diag(m_1,m_2,m_3)$, the mixing $V_\nu$  given in
Eq. (\ref{neutpart}) leads to the following texture for the
neutrino mass matrix:
\begin{eqnarray} M_\nu &&= V_\nu M_\nu^D V^T_\nu \nonumber \\
&&= {\small\left (
\begin{array}{ccccc}
m_1 c^2 + m_2 s^2 &\,\,& (m_2-m_1) cs &\,\,& 0  \\
(m_2-m_1) cs &\,\,&  m_1 s^2 + m_2 c^2 &\,\,& 0  \\
0 &\,\,& 0 &\,\,& m_3
\end{array}
\right )}\,. \label{neumass2} \end{eqnarray}
similarly, taking  $M^D_l=diag(m_e , m_\mu , m_\tau)$, the mixing
matrices $V_{eL}$ and $V_{eR}$ given by the respective
Eqs.~({\ref{clepton}) and (\ref{VeR}), lead to the following
texture for the charged lepton mass matrix,
\begin{eqnarray} M_l&&=V_{eL} M_l^D V^{\dagger}_{eR}\nonumber \\
&& = {\small \left (
\begin{array}{ccccc}
m_e &\,\,& 0 &\,\,& 0  \\
0 &\,\,&   (m_\mu - m_\tau)/2 &\,\,&
- ( m_\mu + m_\tau)/2  \\
0 &\,\,& ( m_\mu + m_\tau)/2 &\,\,& - ( m_\mu - m_\tau)/2
\end{array}
\right )}\,. \nonumber \\ \label{clepmass2} \end{eqnarray}
Extensions of the minimal 3-3-1 model that realize our proposal
must recover the above textures.

In order to assure the feasibility  of this proposal, let us implement an
illustrative scenario which leads to the pattern of lepton mixing presented
in (\ref{neumass2})  and (\ref{clepmass2}). Let us first focus on the matrix
$M_l$ in (\ref{clepmass2}). Perceive that its unique non-diagonal element different
from zero is anti-symmetric. This can be obtained by the scalar triplet $\eta$
once its Yukawa interaction in (\ref{yukawa}) gives rise  to anti-symmetric entries.
The diagonal elements can be generated by the sextet $S$. Then the Yukawa
interactions in (\ref{yukawa}), with specific choice of the Yukawa couplings
$G_{ab}$  and $F_{ab}$, can generate the matrix $M_l$ in (\ref{clepmass2}).

It is the generation of $M_\nu$ given in (\ref{neumass2}) that requires
extension of the minimal model. As we are considering only an illustrative
scenario, one immediate possibility of extension is the inclusion of a second
sextet, $S^{\prime}$, with the Yukawa interaction
\begin{eqnarray}
{\cal L}^Y_\nu =    \frac{1}{2}G^{\prime}_{ab}\overline{ (\Psi_{aL})^c}
S^{\prime *}\Psi_{b_L}.
\label{S2}
\end{eqnarray}
This extension in conjunction with the type II seesaw mechanism
developed in Ref.~\cite{joshi}, and appropriated choice of the
Yukawa couplings, $G^{\prime}_{ab}$, can generate the pattern of
$M_\nu$  given in (\ref{neumass2}).

We finish this section discussing features potentially restrictive for our
proposal due to the form of $O^U$  in (\ref{VUmatrix}), since not only
rare decays are possible, but also
the coupling of the doubly charged vector bilepton to fermions is
maximal (equal to $g/\sqrt{2}$) in some cases, leading to possibly serious
constraints on its mass.
For this reason we discuss now the possible phenomenological
constraints on the proposed extension to the minimal model.

The phenomenology of vector bileptons is well studied in
Ref.~\cite{cuypers}, and here we just update the main results
relevant for this work. First, the rare decay involving the
doubly charged vector bilepton, from the form of $O^U$ in
Eq.~(\ref{VUmatrix}), is the one which mixes the muon and tau,
$\tau^+ \rightarrow \mu^- e^+ e^-$. This decay roughly gives
(assuming $g \sim 0.1$ as for the analysis of $M-\overline{M}$)
$M_{U^{++}}\geq 0.16$ TeV, representing no threat to the model
consistency. Because $O^U_{ee} = 1$ we can also have
contributions to the Bhabha scattering. However, in the realm of
the 3-3-1 model we have to consider the contribution of the $Z'$
as well. The contributions of $U^{++}$ and the $Z'$ have negative
interference terms that blurs the possible effects of the
bilepton, implying that no limits on its mass can be extracted
from the experimental data. Finally, another possible source of
constraint on $U^{++}$ mass would be the anomalous magnetic
moment of the muon, $(g-2)_\mu$. However, due to the
uncertainties related to the hadronic contribution, the deviation
from the SM prediction is not enough to put any severe constraint
on the vector bilepton mass. In this sense, the bound imposed by
consistency of the model remains valid in our suggested scenario.

\section{Concluding Remarks}
\label{sec3}

In this work we  focused on the possibility of the leptonic mixing to help
finding a way out to avoid the $M-\overline{M}$ bound on
$M_{U^{++}}$. By assuming that right-handed and left-handed
charged leptons both present maximal mixing, we arrived at a
pattern of leptonic mixing which accommodates the recent
experimental result in neutrino physics and eliminates the
contribution of the bileptons to the $M-\overline{M}$ conversion\cite{remark}.
Of course we have no knowledge of the actual values for the angles
in the right-handed charged lepton mixing matrix, but the
specific choice we made represents a possible conciliation of
minimal 3-3-1 model with both, neutrino physics and the bound
posed by $M-\overline{M}$ conversion.

For completeness we presented the texture for neutrino and
charged lepton mass matrices in accordance with the pattern of
leptonic mixing used in solving the $M-\overline{M}$ problem. It
serves as a sort of guide to build extensions of the model capable
of recovering our pattern of mixing. Finally we suggested a simple
scenario that could realize our proposal. It is suitable to remark
that the inconsistency raised by the $M-\overline{M}$ bound in
3-3-1 model may be pointing the need of considering charged
leptons a non-diagonal basis.

{\it Acknowledgments.} The authors would like to thank Vicente
Pleitez for the critical reading of the manuscript and for
useful suggestions. This work was supported by Funda\c c\~ao de
Amparo \`a Pesquisa do Estado de S\~ao Paulo (FAPESP)(AG,PSRS) and by
Conselho Nacional de Desenvolvimento Cient\'{\i}fico e
Tecnol\'ogico (CNPq) (CASP)

\end{document}